\documentclass[10pt]{amsart}
\usepackage[utf8]{inputenc}
\usepackage{amsfonts}
\usepackage{amssymb}
\usepackage{amsmath,amsthm}
\usepackage{amscd}
\usepackage{graphics}
\usepackage{cancel}
\usepackage{pdfsync}
\usepackage[british]{babel} 
\usepackage{mathrsfs}
\usepackage{enumitem}
\usepackage{stmaryrd}
\usepackage{wasysym}
\usepackage{latexsym}
\usepackage[colorlinks,linkcolor=blue,citecolor=blue,urlcolor=blue]{hyperref}

\numberwithin{equation}{section}

\newcommand{\beq}{\begin{equation}}
	\newcommand{\eeq}{\end{equation}}
\newcommand{\bea}{\begin{eqnarray}}
	\newcommand{\eea}{\end{eqnarray}}

\newcommand{\bk}{\begin{cases}}
	\newcommand{\ek}{\end{cases}}

\theoremstyle{definition}

\usepackage{enumerate}
\usepackage{float}
\usepackage{mathtools}
\usepackage{afterpage}

\allowdisplaybreaks

\usepackage{bm}

\begin{document}
	
	\title[A note about St\"ackel,  Eisenhart and Haantjes geometries]{A note about St\"ackel,  Eisenhart and Haantjes geometries}
	
	\author{Ond\v{r}ej Kub\r{u}}
	\address{Instituto de Ciencias Matem\'aticas, C/ Nicol\'as Cabrera, No 13--15, 28049 Madrid, Spain}
	%\email{ondrej.kubu@icmat.es}
	
	\author{Piergiulio Tempesta}
	\address{Departamento de F\'{\i}sica Te\'{o}rica, Facultad de Ciencias F\'{\i}sicas, Universidad Complutense de Madrid, 28040 -- Madrid, Spain \\ and Instituto de Ciencias Matem\'aticas, C/ Nicol\'as Cabrera, No 13--15, 28049 Madrid, Spain}
	
	%\date{\today}
	
	\begin{abstract}
		In a recent preprint, we discussed geometric lifts, and in particular the notion of St\"ackel lift. In this note, we wish to clarify several aspects raised in the comment \cite{comment}.
	\end{abstract}
	
	\maketitle
	
	In the recent preprint \cite{KT2025}, we proposed the notion of St\"ackel lift, which, rooted in the classical ideas of St\"ackel, allows us to generate geometric lifts of classical Hamiltonian integrable systems, and, as a particular case, to reproduce  Eisenhart lifts of large classes of systems.
	
	\vspace{2mm}
	
	In the comment \cite{comment}, the notion of St\"ackel lift is interpreted in terms of the classical Jacobi equations. In particular, a system of $n+m$ algebraic equations, regarded as the standard Jacobi separation equations, is proposed as an alternative means of generating the lifts discussed in our preprint. The main conclusion of \cite{comment} is that our concept of a St\"ackel lift can be viewed as an application of the classical Jacobi method, and it is argued that this may cause confusion about the novelty of our approach.
	
	\vspace{1mm}
	
Consequently, to avoid misunderstanding, we wish to clarify several important points.
	
	\vspace{1mm}
	
	As is well known, all methods---both classical and modern---concerning separation of variables, including St\"ackel and Eisenhart geometry, can be interpreted within the framework of Jacobi's ideas, which ultimately reduce to the Jacobi--Sklyanin equations.

	%\vspace{2mm}
	
The analysis in \cite{comment} is carried out within the standard framework of St\"ackel matrices defined   in the configuration space, under the assumption that this is the setting we develop in our preprint.
	
	\vspace{1mm}
	
	In fact, our aim is precisely the opposite. Our purpose is to construct new \textit{physically relevant} Hamiltonian systems which, in general, are separable in non-orthogonal coordinate systems defined in the full phase space, by means of generalized St\"ackel geometry. Notably, in \cite{comment} it was not taken into account that the lifted St\"ackel matrices in our approach depend crucially on the momenta; they are therefore \textit{non-classical}, generalized St\"ackel matrices. In fact, the lifted St\"ackel matrices appearing in preprint \cite{KT2025} have in general  entries $S_{ij}(q,p)$ depending explicitly on momenta, unlike the classical case $S_{ij}(q)$ considered in \cite{comment}. Combining the theory of geometric lifts with generalized St\"ackel geometry is a crucial and novel aspect of our work. The framework proposed in \cite{comment} focusses on a special instance of our  procedure, namely to the case when the St\"ackel matrix is classical, i.e. momentum-independent.
	
	\vspace{1mm}
	
	Beyond this, there is a methodological question that can be raised, regardless of whether standard or generalized St\"ackel matrices are used. One might ask why one would attempt to study an expanded system of $n+m$ Jacobi--Sklyanin equations---which must first be properly constructed, and then solved (possibly, with computer-aided algorithms, as suggested in \cite{comment})---when our procedure achieves the same result directly, by extending the St\"ackel matrix \textit{in the phase space} without solving any system of  equations.  Moreover,   no geometric insight is provided into the form of the proposed $n+m$ equations needed to produce \textit{physically relevant} examples.
	
\vspace{1mm}
		
Concerning the fundamental separation equations,  quite remarkably, from the lifting matrix they can be  easily written in full generality.		In fact, in our approach, these equations are readily derived, \textit{a posteriori}, as a direct consequence of the generalized St\"ackel geometry, rather than being postulated \textit{a priori},  as a system of equations to be properly written and solved.
	
	\vspace{1mm}

We also remark that the abundance of systems that can be constructed via St\"ackel lifts, which obviously reflects the freedom in constructing the lift of a given system, should be regarded as an advantage of our methodology rather than a drawback. The true challenge---and the key difficulty---lies in discovering \textit{meaningful} new systems. This is the reason why a substantial part of our preprint is devoted to demonstrating that, in fact, the systems obtained via the St\"ackel lift can be genuinely significant and interesting.
	
	\vspace{2mm}
	
	From a physical point of view, the relevance of the class of systems we obtain lies in their considerable range of applications: a suitable choice of the lifting matrix can produce Eisenhart-lifted (geodesic) systems, non-geodesic models, lifts of Hamiltonian systems immersed in a magnetic field, and models of gravitational $pp$-waves, which are relevant in modified gravity approaches and Finsler geometries.
	
	\vspace{2mm}
	
	From a mathematical viewpoint, beyond their integrability and separability---which are ensured by the St\"ackel geometry---these systems possess nontrivial symplectic-Haantjes manifolds, which are more general than those associated with the Arnold--Kozlov--Neishtadt systems.
	
	%\vspace{2mm}

	We hope this clarifies the scope and novelty of our approach.

\section*{Acknowledgement}
We wish to thank Professors G. Marmo and G. Tondo for discussions.

\end{document}